\documentclass[10pt, aps, prb, superscriptaddress, noshowkeys, twocolumn,
floatfix, preprintnumbers,showpacs]{revtex4-1}

\usepackage{amsmath, amssymb, amsfonts}
\usepackage{graphicx, color}
\usepackage{times}

\usepackage{graphicx}
\usepackage{bm,color}
\usepackage{times}
\usepackage{color}

\usepackage[breaklinks]{hyperref}
\usepackage{bookmark} 
\hypersetup{colorlinks, citecolor = blue, filecolor = blue, linkcolor =
blue, urlcolor = blue} 
\renewcommand{\Re}{\,\text{Re}}
\newcommand{\bk}{{\boldsymbol{k}}}

\newcommand{\bq}{{\boldsymbol{q}}}
\newcommand{\bK}{{\boldsymbol{K}}}
\newcommand{\bx}{{\boldsymbol{x}}}
\renewcommand{\bm}[1]{\boldsymbol{#1}}

\newcommand{\br}{\bm{r}}
\newcommand{\bR}{\bm{R}}
\newcommand{\brho}{\bm{\rho}}

\allowdisplaybreaks

\begin{document}
\title{Screening and gap generation in bilayer graphene}

\author{T. Stroucken}
\affiliation{Department of Physics and Material Sciences Center,
Philipps University Marburg, Renthof 5, D-35032 Marburg, Germany}

\author{J. H. Gr{\"o}nqvist}
\affiliation{Department of Physics and Material Sciences Center,
Philipps University Marburg, Renthof 5, D-35032 Marburg, Germany}
\affiliation{Department of Physics, \AA bo Akademi University, 20500
Turku, Finland}

\author{S.W. Koch}
\affiliation{Department of Physics and Material Sciences Center,
Philipps University Marburg, Renthof 5, D-35032 Marburg, Germany}

\begin{abstract}

A fully selfconsistent treatment for gap generation and Coulomb screening
in excitonic insulators is presented. The method is based on the
equations of motion for the relevant dynamical variables combined with a
variational approach. Applying the theory for a model system of bilayer graphene, an excitonic groundstate with a gap exceeding 10 meV is predicted. 

\end{abstract}
\date{\today}


\maketitle

\section{Introduction}

With the discovery and rise of graphene and graphene based systems, the
interest in the spontaneous formation of an excitonic groundstate has
been revived.  It has been noticed already in the 1960's that
electron--electron interactions in narrow-gap semiconductors or semi-metals
may lead to an instability of the normal groundstate\cite{Jerome1967,Halperin1968}.
If the noninteracting gap is smaller than the exciton binding energy,
excitons may form spontaneously and the
system is expected to undergo a phase transition into an excitonic state in close
analogy to the BCS superconductor.
In contrast to the BCS superconductor, however, the pairing in the narrow-gap semiconductors occurs
between oppositely charged particles leading to an excitonic insulator
state. Its quasiparticle spectrum exhibits a gap exceeding that of the noninteracting groundstate and it has a characteristic
``Mexican hat''-like shape, displaying
the spontaneously broken symmetry of the interacting groundstate\cite{Jerome1967}

Since its first theoretical proposal, the experimental and theoretical
search for material systems hosting this
interesting state of matter has been a subject of continuous
research\cite{Jerome1967,Halperin1968,Littlewood1996,Butov2004,Eisenstein2004}.
The ideal host combines a small gap with a large exciton binding and
high stability with respect to external perturbations.

As a consequence of its zero-gap single-particle bandstructure, single-layer graphene (SLG) seems a promising candidate for an excitonic
insulator. Due to the linear dispersion SLG exhibits massless, chiral
Dirac Fermions which attracted significant
interest recently. In particular, the
formal analogy between spontaneous exciton
condensation and chiral symmetry breaking in QED has been discussed by
several authors\cite{Semenoff1984,Khveshchenko2009,Wang2010a,Zhang2011,Semenoff2012,Kotov2012}.

The linear dispersion and chiral nature of the quasiparticles
in SLG leads to fundamental differences
in the effects of e--e interactions as compared to the conventional
two-dimensional electron gas (2DEG). Recent
analysis\cite{Gronqvist2012} of the Wannier equation with a linear dispersion shows that exciton binding in SLG requires a minimum
effective Coulomb coupling strength, in general agreement with other theoretical investigations\cite{Pereira2007,Shytov2007a,Shytov2007b,Gamayun2009,
Gamayun2010,Sabio2010b,Wang2010b,Stroucken2011,Drut2009a,Drut2009b,Gonzalez2012,Kotov2012}. 

The actual strength of the Coulombic coupling in SLG is still a subject of debate\cite{Hwang2007,Gamayun2010,Reed2010,Gonzalez2012,Kotov2012,Gonzalez2012,
Gonzalez2012b}.
Combination of the static limit of the standard Lindhard formula with
experimentally measured values for the Fermi velocity predicts values for the Coulombic coupling that are slightly larger than the
critical value, thus allowing for a small but
finite gap.  However, despite intense investigations, so far no
experimental evidence has been presented for the occurrence of a gap in
the SLG spectrum.

Recently, also bilayer graphene (BLG) has attained much interest. 
Combining two graphene sheets to a so called A--B
or Bernal stacked bilayer, the interlayer tunneling changes the linear
dispersion into a quadratic one, preserving the
band degeneracy at the Dirac points and the chiral nature of the
quasiparticles\cite{McCann2006,McCann2007,Nilsson2008}. As such, BLG is a very interesting
system, exhibiting simultaneously similarities and differences in its
excitonic and screening properties as compared
to SLG and the 2DEG. Specifically, the massive quasiparticles lead to bound
excitons for arbitrarily weak Coulomb attraction\cite{Zhang2010,Vafek2010,Nandkishore2010}
and hence, BLG is expected to host an excitonic groundstate. Moreover,
as has been predicted theoretically and demonstrated
experimentally\cite{McCann2006,McCann2007,Stauber2007,Liang2012},
a tunable gap can be introduced by applying a gate bias or
perpendicular electric field and hence, BLG
provides the ideal model system to study exciton condensation.

As has been emphasized by several authors, the predictions for gap
generation and exciton condensation depend sensitively on the model used
for screening. In the existing literature, dynamical screening has
mainly been studied within the standard RPA or one-loop
approximation\cite{Hwang2008,Sensarma2010,Zhang2010,Vafek2010,Gamayun2011,Nandkishore2010}.
Within the standard RPA, the bare bubble polarization describes virtual
transitions across the Fermi level. The prediction for the screening  
depend on the occupation numbers of the involved states, their
dispersion and overlap matrix elements. Unlike in
conventional materials where the bandstructure is fixed by the
crystalline structure, in an excitonic insulator it 
depends on the strength of the Coulomb interaction itself, and hence, the 
screening and gap calculations have to be done selfconsistently.

In this paper, we develop a framework that is  capable to treat gap
generation and screening on equal footing. Our method
is based on the equations of motion for the dynamical variables that are
solved in the static limit. In the regime of
weak Coulomb coupling, our method produces a self-energy correction to
the single particle energies and the standard bare
bubble polarization. In the regime of strong Coulomb coupling, our results predict the opening of a gap in the quasiparticle spectrum that suppresses effective
screening.

We use our method to analyze the excitonic instability in BLG. For this purpose, we adopt a two-band continuum model
that neglects trigonal warping and intervalley interactions. The numerical evaluations
predict a gap of roughly 14 meV for freely suspended BLG. For each
valley degree of freedom, the
corresponding state displays a broken layer symmetry,
leading to a charge polarization. States with a charge polarization
in opposite directions are degenerate. 

The paper is organized as follows: In Sec. \ref{sec:Hamiltonian}, we
present our model Hamiltonian for the BLG system.
In Sec. \ref{sec:screened_gap_eq}, we discuss the screened gap equations
and in Sec. \ref{sec:results} we summarize our numerical results.
A detailed derivation of the screened gap equations is given in the
App. \ref{app:SGE}.

\section{Bilayer model Hamiltonian\label{sec:Hamiltonian}}

\begin{figure}
\centerline{\includegraphics{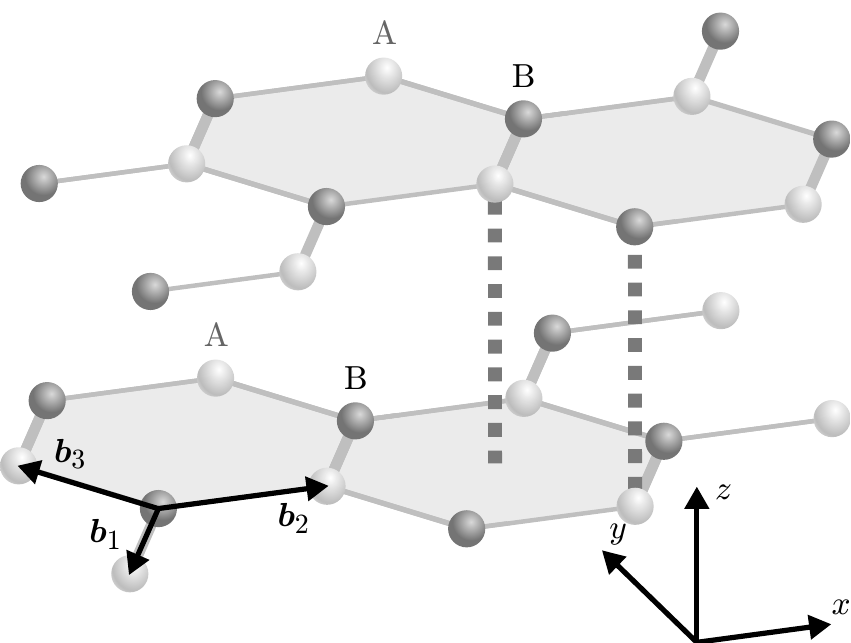}}
\caption{Schematic illustration of the bilayer graphene lattice in
Bernal stacking.
The lattice consists of two honeycomb structures separated by a distance
$L$ and rotated with respect to each other.
Each single layer is characterized by the three vectors ${\bm b}_i$
connecting a carbon atom of the
$A$ layer with its three nearest neighbors.}
\label{Fig:bilayer-lattice}
\end{figure}

\begin{figure}
\centerline{\includegraphics{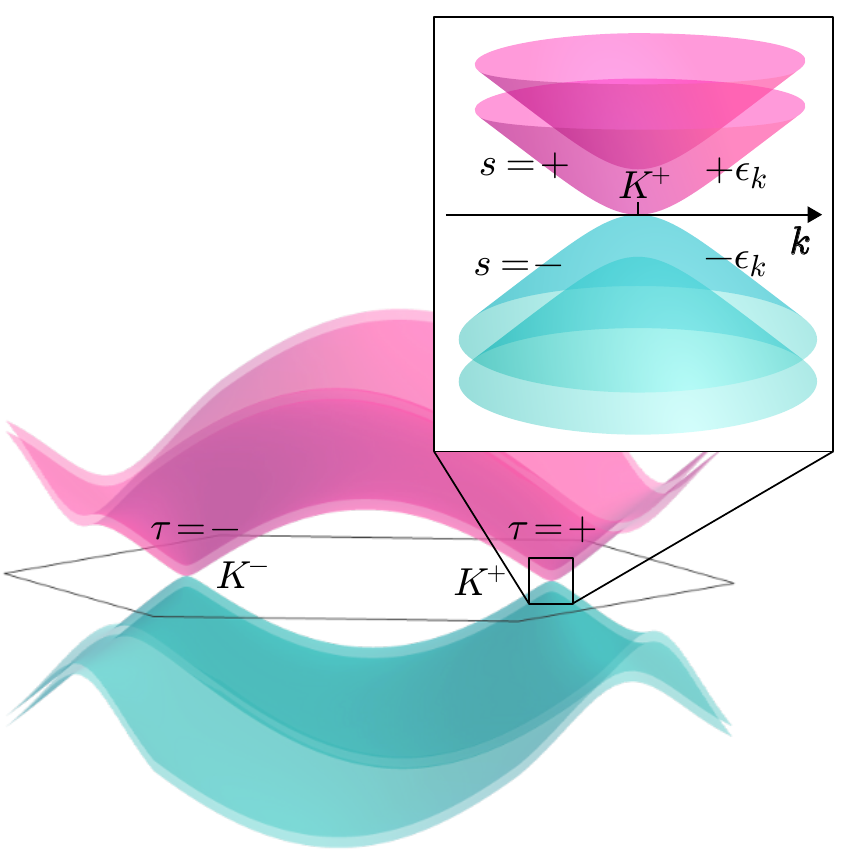}}
\caption{Schematic illustration of the bilayer bandstructure within the tight-binding model. Two of the four bands are degenarate at the Dirac points $K^\pm$, labeled by the valley index
$\tau$. The inset shows a zoom in at the $K^+$ point. The degenerate bands are labeled by the band index $s=\pm1$ and have energies $\epsilon_{s\bk}=s\epsilon_{\bk}$, respectively.} 
\label{Fig:bilayer-bandstructure}
\end{figure}

As a model system, we consider BLG in Bernal stacking, as
shown schematically in Fig. \ref{Fig:bilayer-lattice}. The
two coupled layers have hexagonal carbon lattic structures with sites $A1$, $B1$ and
$A2$, $B2$ located at $z_1$ and
$z_2$ respectively. In the Bernal stacking, the in-plane coordinates of
the $B1$ and $A2$ layers coincide, but those of $A1$
and $B2$ do not. Within a single layer, only nearest neighbor hopping is
taken into account and characterized by the hopping
parameter $\gamma_0=2.8$ eV, which is related to the Fermi velocity by 
$3\gamma_0 b/2\equiv\hbar v_F$ where $b=1.41${\AA} is the
carbon--carbon distance. Interlayer hopping between the sites B1 and A2
is characterized by the hopping parameter
$\gamma_1=0.39$ eV and that between $A1$ and $B2$ by $\gamma_2=0.32$ eV,
respectively. Hence, the free-particle part of the Hamiltonian is given by
\begin{equation}\label{bilayer-h}
 H_0=
   \sum_{\bk}\hat\Psi^\dagger_{\bk}
     \begin{pmatrix} 
             0&\gamma_0f({\bk})&0&\gamma_1\cr
             \gamma_0f^*({\bk})&0&\gamma_2f^*({\bk})&0\cr
             0&\gamma_2f({\bk})&0&\gamma_0f({\bk})\cr
             \gamma_1&0&\gamma_0f^*({\bk})&0
     \end{pmatrix}
   \hat\Psi_{\bk}
\end{equation}
where
$\hat\Psi_{\bk}^\dagger=(a_{1,\bk}^\dagger,b_{1,\bk}^\dagger,a_{2,\bk}^\dagger,b_{2,\bk}^\dagger)$
is a 4-component field operator combining the creation operators in the 4
sublattices and
$f({\bk})=\sum_{i=1}^3e^{i{\bk}\cdot {\bm b}_i}$ reflects the symmetry
of the hexagonal lattice, see Fig.
\ref{Fig:bilayer-lattice}. Within the first Brillouin zone, $f({\bk})$
has two nonequivalent roots
$\bK^\pm = (2\pi/3b,\pm2\pi/3\sqrt{3}b)$ defining the two Dirac points.
Diagonalization of the single-particle Hamiltonian gives four bands, two
of them being degenerate at the Dirac points and two
of them shifted by $\gamma_1$, see Fig. \ref{Fig:bilayer-bandstructure}.
Hence, in the energy range $E<\gamma_1$,
the single-particle part of the effective Hamiltonian
can be described within a two-band approximation, with contributions arising from
regions in the Brillioun zone that are centered around the two Dirac points. 
If we define the wavenumber
with respect to the Dirac points and distinguish the respective Dirac points by the valley index
 $\tau=\pm$, each valley component contributes equally, giving 
\begin{equation}
  H_0=\sum_{\tau, s,\bk}  \epsilon_{s,\bk}c_{\tau,s,\bk}^\dagger
c_{\tau,s,\bk}.
\end{equation}
Here, $c^\dagger_{\tau,s,\bk}$ creates an electron 
with band index $s$ and
wavenumber $\bK^\tau+\bk$. Within the two-band approximation, the symmetric conduction and valence bands
are conveniently labeled by $s=\pm 1$ with $ \epsilon_{s,\bk}=s \epsilon_{\bk}$. The
corresponding Bloch waves consist of components on the interacting $B1$
and $A2$ sublattices localized on the lower and
upper sheet, respectively. A detailed derivation of the projection
procedure is given in App. B. Close to the Dirac
points, the effect of the trigonal warping is negligible and the energy
dispersion of the symmetric bands can be approximated
by the relativistic dispersion
\begin{equation}\label{dispersion}
  \epsilon_{\bk} = \sqrt{\left(mv_F^2\right)^2+(\hbar v_F
k)^2}-mv_F^2
\end{equation}
with the rest mass $m=\gamma_1/(2v_F^2)$ determined by the interlayer
hopping.

As we are interested in collective system properties on a length scale 
larger than the lattice spacing,
we include the Coulomb interaction within the continuum model and
neglect intervalley scattering requiring a momentum
transfer $|\bq|>\frac{4\pi}{3\sqrt{3}b}$. Using a $k$-space representation
for the in-plane coordinates and keeping the full
$z$ dependence, the Coulomb Hamiltonian is 
\begin{eqnarray}
  H_C  &=&
              \frac{1}{2}\frac{e^2}{\epsilon_B} \sum_{\bq\neq 0}
V_0(\bq) \iint dzdz'\,\hat\rho_{\bq}(z) {\rm e}^{-q|z-z'|}
\hat\rho_{-\bq}(z')
                  \nonumber\\
          &=&
\frac{1}{2}\frac{e^2}{\epsilon_B}\sum_{\bq\neq 0}\sum_{ij}V_0^{ij}(\bq)\hat\rho^{i}_{\bq}
\hat\rho^{j}_{-\bq}.
\end{eqnarray}
Here, $V_0(\bq)=2\pi/(qNA)$ is the two-dimensional (2D) Coulomb potential, $NA$ is the
normalization area, and $\epsilon_B$ is the dielectric
constant of the substrate. Furthermore,
\begin{eqnarray}
     \hat\rho_{\bq}(z)  &=&
           \sum_{{\bk}}\sum_{\tau,ss'}
                                                W_{\tau,ss'}(\bk,\bq,z)
                                               c^\dagger_{\tau,s,\bk-\bq}c_{\tau,s',\bk}
                                                                 \nonumber\\
        &\equiv&
                          \sum_{i=1,2}\hat\rho^i_{\bq}f({\bq},z-z_i)
\label{defchargedensity}
\end{eqnarray}
is the effective density operator  projected onto the lowest bands,
\begin{align*}
        W_{\tau,ss'}(\bk,\bq,z) = \frac{1}{2} \bigl[ {\rm
e}^{-i\tau(\phi_{\bk}-\phi_{\bk-\bq})} &f({\bq},z-z_1)\\
                     + ss' {\rm e}^{i\tau(\phi_{\bk}-\phi_{\bk-\bq})}
&f({\bq},z-z_2) \bigr]
\end{align*}
is the wave function overlap, and
$f(\bq,z)=\int_{e.c.} d^2\rho {\rm e}^{-i\bq\cdot\brho}|\phi(\br)|^2$
is the in-plane Fourier transform of the electron density corresponding
to the atomic orbitals.
Within the tight-binding (TB) approximation, the total charge density can be divided
into the contributions
$\rho^i$, located in layer $i$. The Coulomb potential between the
carriers in sheet $i$ and $j$ is
\begin{align}
         V_0^{ij}(\bq)
             &= V_0(\bq) \iint \!\! dz dz'\,f(\bq,z-z_i){\rm
e}^{-q|z-z'|}f(-\bq,z'-z_j)\nonumber\\
             &= V_0(\bq)H(q,|z_i-z_j|) \, .
\end{align}
As shown in  App. \ref{app:scales}, $H(q,L)\equiv\tilde H(qd,L/d)$ depends on the
effective thickness $d$ of a single graphene
sheet and the interlayer spacing $L$.
These intrinsic length scales should be compared to the excitonic Bohr
radius
\[
a_0=\frac{\hbar^2\epsilon_B}{me^2}=
3\frac{\gamma_0}{\gamma_1}\frac{b}{\alpha}
\]
where $\alpha=e^2/(\epsilon_B\hbar v_F)$ is the effective fine structure
constant.
If $d/a_0<<1$, finite size effects on the excitonic properties can be
neglected and the system can be
considered as effectively 2D.
Within this limit, the Hamiltonian scales strictly with the exciton Bohr
radius and depends on the parameter combination $\alpha$ characterizing
the importance of relativistic effects. In the nonrelativistic limit
$\alpha\ll1$, the BLG dispersion is approximately quadratic and the only
relevant energy unit is
the exciton Rydberg
\[E_0=\frac{m e^4}{\epsilon_B^2\hbar^2}=\frac{1}{2}\alpha^2\gamma_1.\]

\section{Screened gap equations}\label{sec:screened_gap_eq}

To determine the groundstate of our model Hamiltonian, we extend the
methods derived in \cite{Stroucken2011} for SLG to treat gap generation and screening effects on equal footing.
In the appendix, we present the technical details of our method that is valid for arbitrary narrow gap semiconductors and semimetals.
Here, we only summarize the main derivation steps adapted specifically to the BLG model Hamiltonian.

In general, the effect of screening is described by the polarization
function,
describing the charge density induced by a density fluctuation. 
In an effectively 2D system, the charge density is  homogeneous with
respect to the in-plane coordinates but localized
in the $z$ direction. For BLG, the charge localization invloves both
layers leading to a matrix-like
Coulomb potential $V^{ij}$ describing the interaction between particles
located in the sheets $i$ and $j$, respectively. As
the polarization results from the induced charge densities within the
layers, it also has a $2\times2$ matrix structure:
\begin{eqnarray}
           \Pi(\omega,\bq,z,z') &=&
-\frac{i}{\hbar}\int_0^\infty dt{\rm e}^{i\omega t}\left<\left[\hat\rho_{\bq}(z,t),\hat\rho_{-\bq}(z)\right]\right> \nonumber\\
&=&\sum_{i,j}f(\bq,z-z_i)
\Pi^{ij}(\omega,\bq)f(-\bq,z'-z_j) \, .\nonumber\\
\label{densdensresponse}
\end{eqnarray}
The screened potential obeys the integral equation
\begin{align*}
                 &V_s(\omega,\bq,z,z') = V_0(\bq,z-z')\\
                 &+ \frac{e^2}{\epsilon_{\rm B}}\iint  dz_1dz_2
V_0(\bq,z-z_1) \Pi(\omega,\bq,z_1,z_2)V_s(\omega,\bq,z_2,z').
\end{align*}
Defining the screened interaction matrix
\begin{eqnarray}
              V_s^{ij}(\omega,\bq) = \iint dzdz'f_{\bq}(z-z_i)V_s(\omega,\bq,z,z')f_{-\bq}(z'-z_j),\nonumber\\
\end{eqnarray}
the integral equation reduces to a much simpler matrix equation with the
solution
\begin{eqnarray}
          {\underline{\underline{V_s}}}(\omega,\bq)   & = &
                    \left(\underline{\underline{I\!\!
I}}-\frac{e^2}{\epsilon_{\rm
B}}{\underline{\underline{V}}}_0(\bq){\underline{\underline{\Pi}}}(\omega,                                                                  
                    \bq)\right)^{-1}
                  {\underline{\underline{V}}}_0(\bq)\nonumber\\
                   & \equiv &
{\underline{\underline{\epsilon}}}(\omega,\bq)^{-1}{\underline{\underline{V}}}_0(\bq).
\end{eqnarray}
If the influence of the finite layer separation $L$ can be neglected, 
the interaction is independent of the layer indices and one obtains the standard 2D result
\begin{equation}
V_s^{2D}(\omega,\bq)=\frac{V_0(\bq)}{1-\frac{e^2}{\epsilon_{\rm B}}V_0(\bq)\Pi^{2D}(\omega,\bq)}
\end{equation}
with $\Pi^{2D}(\omega,\bq)=\sum_{ij}\Pi^{ij}(\omega,\bq)$.
The components of the dynamical polarization ${\underline{\underline{\Pi}}}(\omega,\bq)$ 
can be calculated from the microscopic equations of motion for  the operator combinations
$c^\dagger_{\tau,s,\bk-\bq}c_{\tau,s',\bk}$ that constitute the density operator.

Due to its specific mathematical
structure as continued fraction, the polarization function can be
divided into a dominant part and small corrections
that can be treated perturbatively. In a weakly interacting system, a convenient choice for the dominant part is 
the noninteracting suceptibility, i.e. expectation values are taken with respect to the noninteracting groundstate.
However,  in a strongly interacting system, corrections due to the Coulomb interaction modify the system 
groundstate properties  and cannot be treated perturbatively. 
Instead, the expectation value on the RHS of Eq. \ref{densdensresponse} 
has to be taken with respect to the interacting groundstate.
Since the interacting groundstate depends on the strength of the Coulomb interaction that is in turn
limited by screening effects, this constitutes a self-consistency problem.

In the absence of external perturbations, the system is homogeneous with
respect to the in-plane coordinates
and only single-particle operator combinations with $\bk=\bk'$, $\tau=\tau'$ give
nonzero expectation values.
Hence,  within a mean-field approximation, the unperturbed Hamiltonian
consists of two independent contributions from each valley component
which are related by the parity transformation or
charge conjugation:
$W_{+,ss'}(\bk,\bq,z)=W_{-,ss'}(-\bk,-\bq,-z)=W^*_{-,ss'}(\bk,\bq,z)$.
In the following, we suppress the valley index and perform explicit
calculations for the $\bK^+$ valley.
As dynamical quantities, we use the microscopic intraband occupation
numbers
$f_{s,\bk} = \langle  c^\dagger_{s\bk}c_{s\bk}\rangle $ and the
interband coherences
$P_{\bk}= \langle  c^\dagger_{-\bk}c_{+\bk}\rangle $.

As shown in App. \ref{app:SGE}, within the time dependent Hartree-Fock approximation, 
the equations of motion for all possible single-particle operator combinations $ c^\dagger_{s\bk-\bq}c_{s'\bk}$
can be derived from the effective mean-field Hamilonian
\begin{eqnarray}
       \hat{H}^{\text{MF}} \!\!
                      &=&
                            \sum_{s,\bk} \Sigma_{s,\bk}
c^\dagger_{s\bk}c_{s\bk}
                                 - \Omega_{\bk} c^\dagger_{+.\bk}c_{,-,\bk}-
                       \Omega^*_{\bk} c^\dagger_{-,\bk}c_{+,\bk},\nonumber\\
\label{HMF}
\end{eqnarray}
where
\begin{widetext}
\begin{eqnarray}
            \Sigma_{s,\bk}  &=&
                               \frac{\delta  \langle  H\rangle }{\delta
f_{s\bk}}
=
s\left(\epsilon_{\bk}+\frac{e^2}{2\epsilon_{\rm B}}\sum_{\bk'}V_s^{\rm
ND}(\bk-\bk'){\rm cos}(2(\phi_{\bk}-\phi_{\bk'}))
                             \left(f_{-,\bk'}-f_{+,\bk'}\right)\right.
                                          \nonumber\\
                                     &-&
                            \left.i\frac{e^2}{2\epsilon_{\rm
B}}\sum_{\bk'}V_s^{\rm ND}(\bk-\bk'){\rm sin}(2(\phi_{\bk}-\phi_{\bk'}))
                            \left(P_{\bk'}-P^*_{\bk'}\right)\right)
                            -\frac{e^2}{2\epsilon_{\rm
B}}\sum_{\bk'}V_s^{\rm D}(\bk-\bk')\left(f_{-,\bk'}+f_{+,\bk'}\right)
\label{def:sigma_s}
\end{eqnarray}
are the renormalized single-particle energies, and
\begin{eqnarray}
                \Omega_{\bk}    &=&
                        -\frac{\delta  \langle  H\rangle  }{\delta
P^*_{\bk}}
                        = i\frac{e^2}{2\epsilon_{\rm B}}\sum_{\bk'}V_s^{\rm
ND}(\bk-\bk'){\rm sin}(2(\phi_{\bk}-\phi_{\bk'}))
                         \left(f_{-,\bk'}+f_{+,\bk'}\right)
                                 \nonumber\\
                            &+&
                          \frac{e^2}{2\epsilon_{\rm B}}\sum_{\bk'}V_s^{\rm
D}(\bk-\bk')
                           \left(P_{\bk'}+P^*_{\bk'}\right)
                          -\frac{e^2}{2\epsilon_{\rm
B}}\sum_{\bk'}V_s^{\rm ND}(\bk-\bk'){\rm cos}(2(\phi_{\bk}-\phi_{\bk'}))
                             \left(P_{\bk'}-P^*_{\bk'}\right)
\end{eqnarray}
\end{widetext}
is the internal field. The energy renormalizations and the internal field result from the exchange contributions of 
the two-particle interaction and are evaluated with the screened Coulomb potential.
Here, we used the notation  $V_s^{\rm D}=V_s^{11}=V_s^{22}$ for the intralayer and $V_s^{\rm
ND}=V_s^{12}=V_s^{21}$ for the interlayer Coulomb interaction, respectively.
 The renormalized single-particle energies in Eq. \ref{def:sigma_s}
contain a constant, band independent contribution that corresponds to a
shift of the Fermi energy which can be dropped from the equations.

Defining the Liouville operator by
\begin{eqnarray}
{\cal L}_{ss'\sigma\sigma'}(\bk,\bq)&=&\delta_{s\sigma}\delta_{s'\sigma'}\left(\Sigma_{s',\bk}-\Sigma_{s,\bk-\bq}\right)\nonumber\\
&-&\delta_{s\sigma}\left(\delta_{s'+}\delta_{\sigma'-}\Omega_{\bk}+\delta_{s'-}\delta_{\sigma'+}\Omega^*_{\bk}\right)\nonumber\\
&+&\delta_{s'\sigma'}\left(\delta_{s+}\delta_{\sigma-}\Omega^*_{\bk-\bq}+\delta_{s-}\delta_{\sigma+}\Omega_{\bk-\bq}\right),
\end{eqnarray}
the time evolution of the density operator is given by
\begin{eqnarray}
\hat\rho^i_{\bq}(t)&=&\sum_{ss'\sigma\sigma'}\sum_{\bk}W^i_{ss'}(\bk,\bq){\rm e}^{-i{\cal L}_{ss'\sigma\sigma'}(\bk,\bq)t/\hbar}
c^\dagger_{\sigma\bk-\bq}c_{\sigma'\bk}\nonumber\\
\end{eqnarray}
such that 
\begin{eqnarray}
\Pi^{ij}(\omega,\bq)&=&{\cal N}\sum_{ss'\sigma\sigma'}\sum_{\bk}W^i_{ss'}(\bk,\bq)\left[\hbar\omega {\cal E}-{\cal L}(\bk,\bq)\right]^{-1}_{ss'\sigma\sigma'}\nonumber\\
&\times&
\left<\left[c^\dagger_{\sigma\bk-\bq}c_{\sigma'\bk},\hat\rho^j_{-\bq}\right]\right>.
\label{PIij}
\end{eqnarray}
Here, ${\cal E}$ is the unity matrix and we included a degeneracy factor ${\cal N}=4$ to account for the spin and valley degrees of freedom.
The expectation value of the equal time commutator  can be evaluated, giving
\begin{eqnarray}
\left<\left[c^\dagger_{\sigma\bk-\bq}c_{\sigma'\bk},\hat\rho^j_{-\bq}\right]\right>
=W_{\sigma\sigma'}^{j*}(\bk,\bq)
\left(f_{\sigma,\bk-\bq}-f_{\sigma',\bk}\right)
\nonumber\\
+\delta_{\sigma-}W_{+\sigma'}^{j*}(\bk,\bq)P_{\bk-\bq}
+\delta_{\sigma+}W_{-\sigma'}^{j*}(\bk,\bq)P^*_{\bk-\bq}\nonumber\\
-\delta_{\sigma'+}W_{\sigma-}^{j*}(\bk,\bq)P_{\bk}
-\delta_{\sigma'-}W_{\sigma+}^{j*}(\bk,\bq)P^*_{\bk}
.\nonumber\\
\end{eqnarray}

In general, the density-density response function 
describes the polarizability due to real and virtual transitions, exhibiting resonances at
the transition energy between the involved states. For the interacting system, the
poles of the polarization occur at $\omega=\pm(E_{s,\bk-\bq}-E_{s',\bk})$ where
\begin{equation}
                E_{s,\bk} = s \sqrt{\Sigma_\bk^2 + |\Omega_\bk|^2}\equiv
s E_{\bk}
 \label{eq:newband}
\end{equation}
and $\Sigma_{\bk}=(\Sigma_{+,\bk}-\Sigma_{-,\bk})/2$ is the spectrum of the mean-field Hamiltonian defined in Eq. \ref{HMF}.
This Hamiltonian can be diagonalized 
by the Bogoliubov transformation with the Bogoliubov vacuum as groundstate.

Using the Bogoliubov representation, the mean-field Hamiltonian and the Liouville operator are diagonal in the band indices
and the transition probability depends on the
occupation numbers of the involved states and overlap
matrix elements. Due to the Pauli exclusion,
intraband transitions require partially filled bands and  do not
contribute to the groundstate polarization.
Hence, within the Bogoliubov picture, the groundstate polarization is solely due to
interband transitions. In a gapped system, 
these transitions require a finite transition energy $\hbar\omega\simeq E_{gap}$ and do not contribute to the static limit
of the groundstate polarization. Moreover, due to the orthogonality of the
Bloch waves with equal wavenumbers, the long-wavelength limit $\bq\rightarrow 0$ of the interband contributions
vanishes exactly in a gapped system and the long-ranged part of the static Coulomb interaction is essentially unscreened.
However, if the system is ungapped, virtual interband transitions
are not ruled out by energy conservation, leading to a finite groundstate polarization even in the static limit.
Thus, the occurrence of a finite gap will reduce the effective screening of the long-ranged part of the static Coulomb interaction, and thus modifies the collective system properties significantly.

As can be recognized from Eq. \ref{eq:newband}, the Bogoliubov spectrum
shows a gap if either
the renormalized single-particle energy, or the internal field is
nonzero at the Dirac points.
For symmetry reasons, a nonvanishing energy renormalization at $\bk=0$
requires anisotropic distributions, while a finite
internal field can also be introduced by a real, isotropic distribution
for the interband coherences.
Physically, the real part of the macroscopic interband coherence
$P=\sum_{\bk}P_{\bk}$ corresponds to a layer
polarization, which can be seen from the average charge density given by
\begin{align*}
               &\rho_{\bq=0}(z) = \frac{1}{2}
                       \sum_{s\bk} f_{s,\bk} \left(
f(\bq=0,z-z_1)+f(\bq=0,z-z_2) \right) \\
                 &+ \sum_{\bk} \Re\left[ P_{\bk} \right] \left(
f(\bq=0,z-z_1) - f(\bq=0,z-z_2) \right).
\end{align*}
Hence, a gap can be achieved e.g. by a static or time dependent external
field that induces either an anisotropy, e.g. by photon absorption, or
a layer polarization, e.g. by a static external field.

Here, we investigate the possibility for a spontaneous gap by
exciton formation.  A gapped state will emerge spontaneously if  a 
symmetry breaking lowers the total energy of the interacting system.

Starting from an arbitrary mean-field state,
an infinitesimal variation of the occupation numbers
and the interband coherence yields an energy shift
\begin{eqnarray}
              \delta\langle  H\rangle   &=&
                              \sum_{s\bk}\Sigma_{s,\bk}\delta
f_{s,\bk}-\sum_{\bk}\left(\Omega_{\bk}\delta
P^*_{\bk}+\Omega^*_{\bk}\delta P_{\bk}\right).
\end{eqnarray}
Assuming that the interacting groundstate
is adiabatically connected to the noninteracting one, the constraint of
a stationary solution
implies (for a detailed derivation see App. A)
\begin{eqnarray}
           f_{+,\bk}&=&1-f_{-,\bk}\equiv f_{\bk}\\
           |P_{\bk}|^2&=&f_{\bk}(1-f_{\bk}),
\end{eqnarray}
giving
\begin{eqnarray}
              \delta\langle  H\rangle   &=&
              \sum_{\bk}\left(2\Sigma_{\bk}P_{\bk}-(1-2f_{\bk})\Omega_{\bk}\right)\delta P^*_{\bk}+c.c.,
\end{eqnarray}
which is negative if the Wannier equation has bound-state solutions\cite{Jerome1967,Gronqvist2012}.
The groundstate populations are determined by the variational principle $\delta<H>=0$,
 giving the algebraic relations
\begin{eqnarray}
           |P_{\bk}|&=&\frac{1}{2}\frac{|\Omega_{\bk}|}{E_{\bk}}
\label{defP}\\
           f_{\bk}&=& \frac{1}{2}\left(1-\frac{\Sigma_{\bk}}{E_{\bk}}
\right).\label{deff}
\end{eqnarray}
 Assuming real, isotropic
distributions, insertion of these relations into the definitions
for the internal field and renormalized energies gives the set of
coupled integral equations:
\begin{align}
             \Omega_{\bk}=& \frac{e^2}{2\epsilon_{\rm
B}}\sum_{\bk'}V_s^{\rm D}(|\bk-\bk'|)
              \frac{\Omega_{\bk'}}{E_{\bk'}}\label{gapomega}\\
              \Sigma_{\bk}=& \epsilon_{\bk}+
              \frac{e^2}{2\epsilon_{\rm B}}\sum_{\bk'}V_s^{\rm
ND}(|\bk-\bk'|){\rm cos}2(\phi_{\bk}-\phi_{\bk'})
              \frac{\Sigma_{\bk'}}{E_{\bk'}}.\label{gapsigma}
\end{align}
The gap equations contain the statically screened Coulomb potential and must be solved selfconsistently with
the static limit of the polarization function $\Pi^{ij}(\omega=0,\bq)$. 

The gap equation Eq. \ref{gapomega} always has the trivial solution
$P_{\bk}=\Omega_{\bk}\equiv 0$
corresponding to the noninteracting groundstate. If this trivial
solution is unique, the interacting system has the same groundstate as the noninteracting one,
i.e. we are in the limit of weak Coulomb coupling. The corresponding mean-field
Hamiltonian
is diagonal within the noninteracting picture, and hence, the Bogoliubov
and noninteracting representations coincide.

The Bogoliubov quasiparticle spectrum is then equivalent to the
renormalized single-particle energy:
\begin{equation}
          \Sigma_{\bk}= \epsilon_{\bk}+
                                  \frac{e^2}{2\epsilon_{\rm
B}}\sum_{\bk'}V_s^{\rm ND}(|\bk-\bk'|){\rm cos}2(\phi_{\bk}-\phi_{\bk'}),
\end{equation}
and contains the renormalization by the filled valence band.
Neglecting the effects of the finite thickness of the bilayer (i.e.
putting $H(q,L)\approx 1$),
this single-particle renormalization corresponds to that of the renormalization group (RG) 
approach\cite{Gonzalez1994,Vafek2010}. 
The corresponding noninteracting polarization function is given by
the standard Lindhard formula with self energy corrections \cite{Hwang2008}.
\begin{eqnarray}
\Pi^{2D}_0(\omega,\bq)&=&\sum_{ij}\Pi^{ij}_0(\omega,\bq)\nonumber\\
&=&{\cal N}\sum_{\bk}|W_{+-}(\bk,\bq)|^2
\left(\frac{1}{\hbar\omega-\left(\Sigma_{\bk}+\Sigma_{\bk-\bq}\right)}\right.\nonumber\\
&-&\left.\frac{1}{\hbar\omega+\left(\Sigma_{\bk}+\Sigma_{\bk-\bq}\right)}\right).
\end{eqnarray}

If the gap equation  has a  nontrivial solution, it
predicts a gap $|\Omega_{\bk=0}|$ in the interacting quasiparticle
spectrum.
The corresponding state is energetically 
below the weakly coupled state and we shall refer to this state as strongly coupled (ground-) state.
The gapped dispersion reduces the long wavelength limit of the static polarizability thus leading to a further increase of the predicted gap. 
Hence, though the noninteracting polarizability may be used as criterion for the occurence of an excitonic instability
of the groundstate, to estimate the size of the gap screening and gap generation must be treated selfconsistently. 

The strongly coupled state has a macroscopic layer polarization
$P=\sum_{\bk}\Omega_{\bk}/2E_{\bk}$.
As can be recognized from Eq. \ref{gapomega}, solutions with opposite
layer polarizations are degenerate.
As Eq.~\ref{gapomega} is valid for both Dirac points, states where
charge-polarization contributions
from the two valley components are aligned or anti parallel are also
degenerate such that a gapped state not
necessarily exhibits a physical charge polarization. However,  this
degeneracy might be lifted by intervalley
interactions which we neglected here.

\section{Numerical results} \label{sec:results}

The screened gap equations constitute a rather complicated set of
integral equations that can be solved iteratively.
In a first step, we evaluate the Lindhard formula in the noninteracting
limit and insert this result into the gap equations.
For any fixed polarization function, the screened gap equations are then
solved by iterating candidates for the solution.
After convergence, $f_{\bk}$ and $P_{\bk}$ are calculated from the
solution of the gap equations
and inserted into the generalized Lindhard formula Eq.~\ref{PIij}, which is used to
modify the Coulomb interaction. This
procedure is repeated until overall convergence is achieved.
All calculations have been performed in the limit of vanishing
interlayer spacing $L\rightarrow 0$, which we verified to
give correct results within the numerical accuracy for realistic values
of $L$.

\begin{figure}
\includegraphics{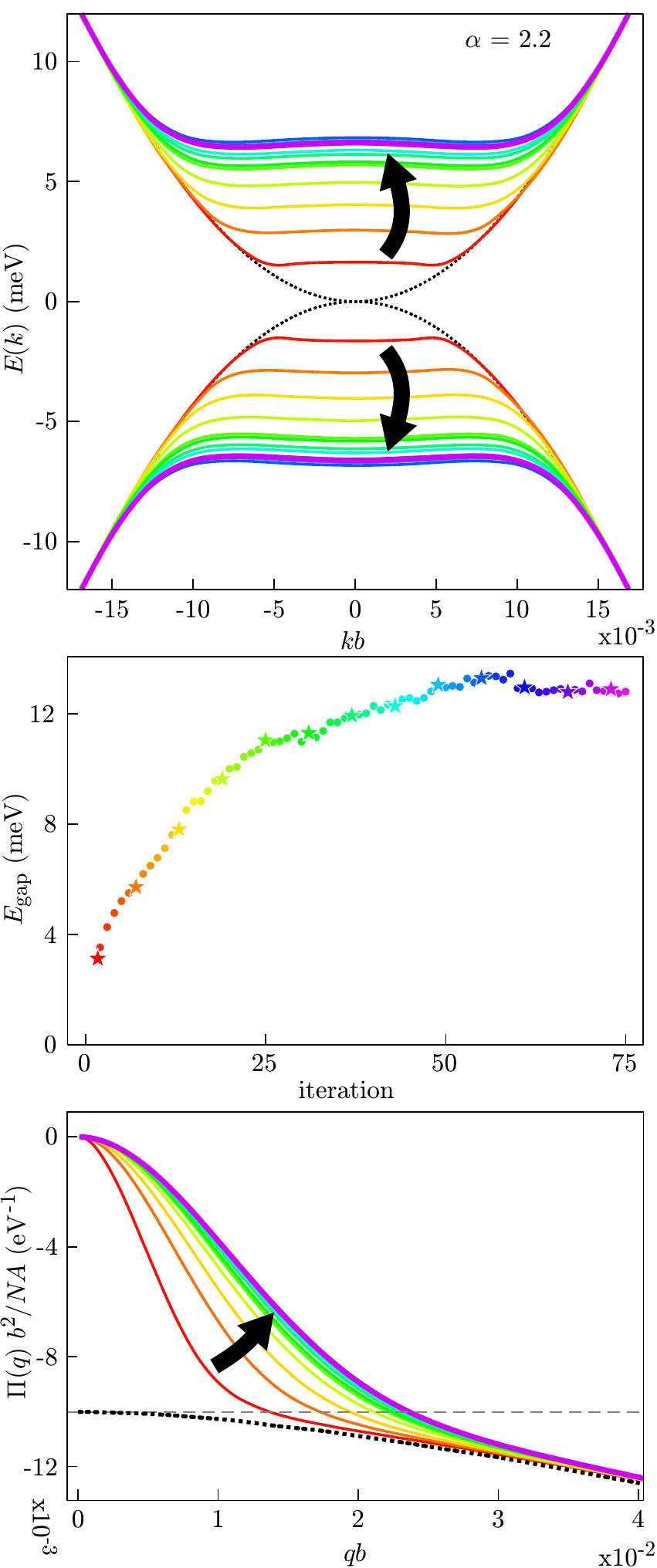}
\caption{Iteration steps in the self-consistent solution of the screened
gap equations \eqref{gapomega}, \eqref{gapsigma}, with the screening
\eqref{PIij}.
In the top panel, the quasiparticle dispersion \eqref{eq:newband} is
shown, in the middle panel, the value of the gap at each step is
plotted, and in the bottom panel, the polarizability is presented.
Here, the screening in the unpopulated tight-binding bands using the
full relativitic disperion is indicated by the dotted line, and the
dashed line shows the constant value obtained in the noninteracting,
nonrelativistic limit.}
\label{iterations}
\end{figure}

Fig. \ref{iterations} shows the resulting spectra and polarization as
functions of $kb$ in the vicinity of the Dirac points. In the calculations, we used
$\alpha=2.2$, which is slightly below the nominal value for the coupling
strength in vacuum.
The individual lines refer to different numbers of iterations and the thick
black arrows indicate the directions of convergence for increasing iteration numbers.
The initial input values, corresponding to the noninteracting groundstate,
are marked by the dotted lines. As can be recognized from the lowest
panel of Fig. \ref{iterations},
the noninteracting polarizability overestimates the effects of
screening, particularly in the long-wavelength limit $q\rightarrow 0$.
As a result of the ungapped dispersion, virtual electron--hole pairs can
be created practically without any cost of energy,
leading to a very efficient intrinsic screening. In the long-wavelength
limit, the noninteracting polarizability
approaches the constant static polarization that is obtained if a purely
quadratic dispersion is
assumed\cite{Hwang2008,Nandkishore2010}. This result, indicated by the dashed
line in the right panel of Fig. \ref{iterations}, turns into a linear curve for
larger wavenumbers, similar to the
noninteracting polarizability of SLG and
characteristic for a linear dispersion\cite{Hwang2007}.

Since massive quasiparticles are sensitive to arbitrarily weak
interactions, we find a nontrivial solution of the gap
equations even if the Coulomb potential is screened by the
noninteracting polarization function.
The resulting quasiparticle spectrum is shown by the red line in the
upper panel of Fig. \ref{iterations} and exhibits a small but
finite gap at the Dirac point. The opening of a gap immediately
suppresses virtual transitions by means of energy conservation,
leading to a practically unscreened long ranged tail of the Coulomb
interaction. In the next step of iteration, this suppression
of the screening enhances the gap, suppressing the screening and further
enhancing the gap, and so on.
Overall convergence is achieved after ca. 70 iterations, as shown in the
middle panel of Fig. \ref{iterations}, where the gap is plotted against the
number of iteration steps. Further iterations give values for the gap
oscillating around its mean value by $\pm 5\%$, indicating the
accuracy of our procedure.

The resulting spectrum exhibits a ``Mexican hat'' shape with band
extrema slightly shifted away from the Dirac point,
typical for an excitonic insulator\cite{Jerome1967}. The  $k$-values denoted by $k_X$, at
which the band extrema occur, can be associated with a
characteristic length scale which can be interpreted as excitonic correlation
length. Between the band extrema, the dispersion is almost
flat, while it approaches the noninteracting dispersion very rapidly for
larger $k$-values. The size of the gap is approximately $E_{gap} \approx
12$ meV,
which is well below the energy of the remote bands falling well within the region
where the quadratic approximation is valid.

\begin{figure*}
\includegraphics{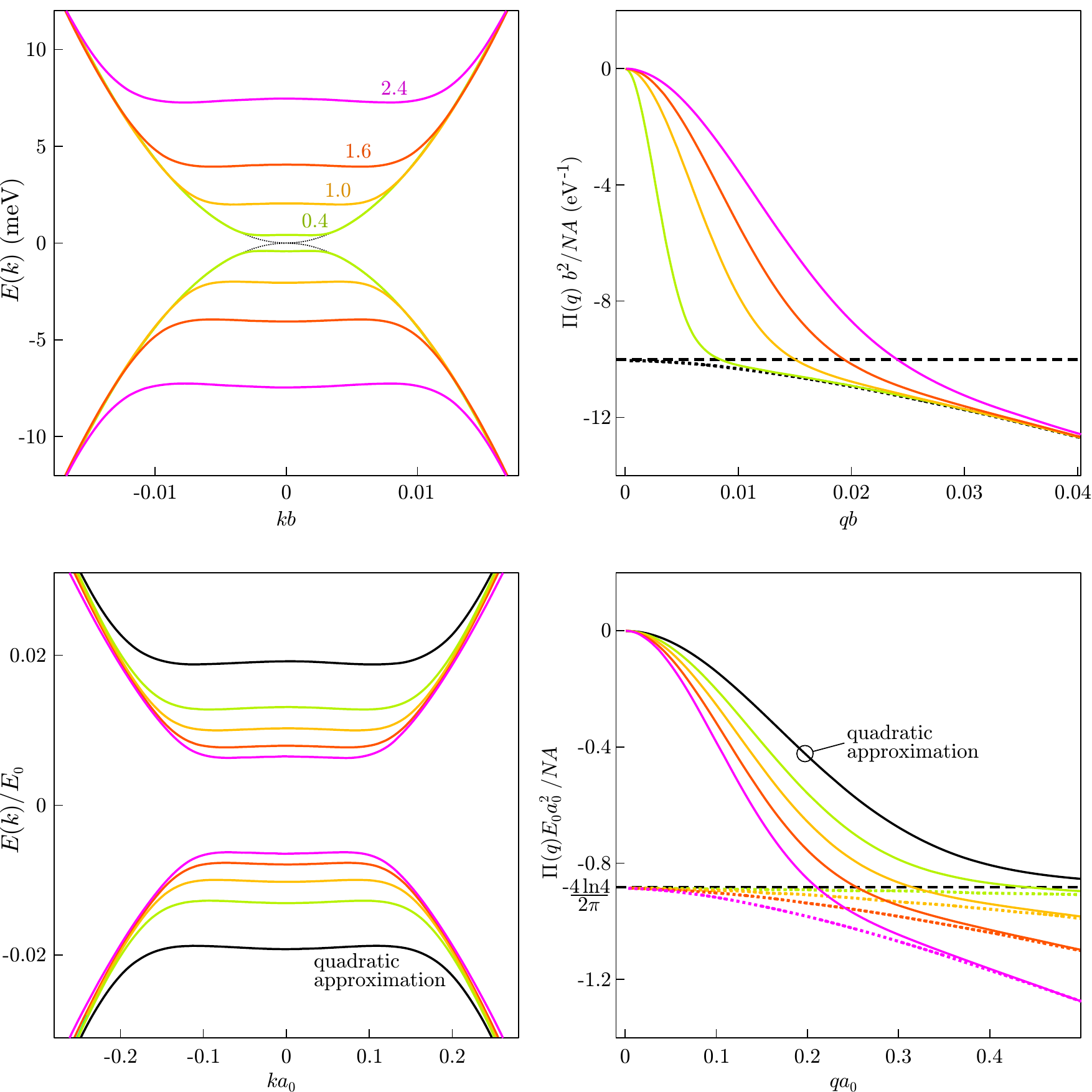}
\caption{Bogoliubov dispersions and polarizabilities as obtained from
the selfconsistently solved screened gap equations.
The lines of different color correspond to the respective solutions for the values $\alpha =$ 2.4, 1.6, 1.0, and 0.4. 
The upper two panels show the solutions in absolute units, and the lower
two in excitonic units, respectively. The panels on the right
show the polarizabilities with zero populations as dotted lines. The dashed line shows the constant value obtained for zero populations in the quadratic
approximation.}
\label{dispersion_and_polarization}
\end{figure*} 

To analyze the dependence on the coupling strength, we plot in Fig.
\ref{dispersion_and_polarization} the Bogoliubov
spectra and the polarizabilities for various values of $\alpha$ in absolute
and excitonic units. With increasing coupling strength, the
characteristic wave number $k_X$ increases, corresponding to shorter
excitonic correlation lengths.
In the lower panel of Fig.  \ref{dispersion_and_polarization}, the same
spectra are shown in
excitonic units. Within the quadratic approximation, all energies scale
with the excitonic energy unit  and all curves merge into
a universal one, indicated by the black line. Despite the small size of
the gap, using the full relativistic dispersion shows
clear deviations from this universal curve, with deviations increasing
with increasing coupling strength.
\begin{figure}
\includegraphics{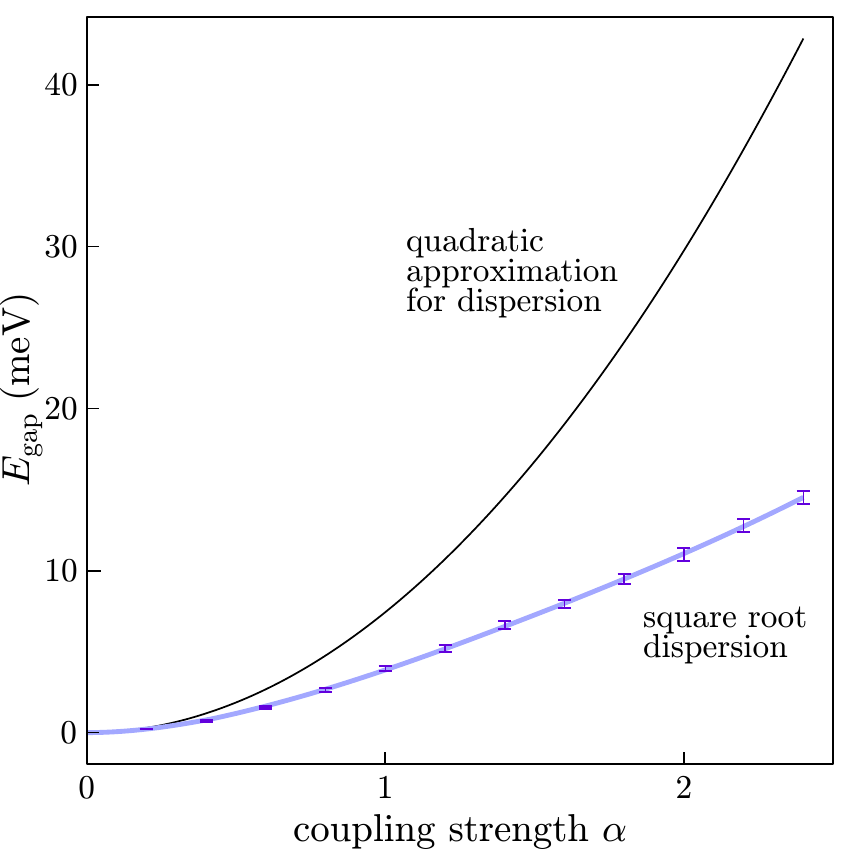}
\caption{Quasiparticle gap vs. coupling strength as obtained from the
selfconsistently solved screened gap equations. The thicker line shows
the values obtained with the full relativisic dispersion
\eqref{dispersion}, whereas the the thinner line indicates the results with
a quadratic
approximation for the dispersion.}
\label{gap_vs_alpha}
\end{figure}
The cause of these deviations can be found in the interplay between the two
different
length scales set by the Compton wavelength and the screening length of
the
massive quasiparticles. The Compton wavelength
$\lambda_c=\hbar/mv_F=3\gamma_0 b/\gamma_1$ involves only single-particle properties and $k\lambda_c\sim 1$ marks the crossover from
the nonrelativistic quadratic 
into the relativistic linear dispersion.
The screening length determines the range of the Coulomb interaction.
Only if the screening length is
large compared to the Compton wavelength, the Coulomb integrals converge
within a wavenumber range where
relativistic corrections to the dispersion can be neglected.
Since the condition of simultaneous energy and momentum
conservation is much less restrictive for the linear dispersion than for
the quadratic one,
the major effect of these corrections is an enhancement of screening
effects.

The inverse screening length follows from the noninteracting polarization
function via $\kappa=-2\pi e^2/\epsilon_B \lim_{q\rightarrow
0}\Pi_0(\omega=0,\bq)$ and is proportional to
$e^2/\epsilon_B$, i.e. to the coupling constant. The long-wavelength
limit of the noninteracting polarizability
can be obtained analytically by assuming a purely
quadratic dispersion, giving $\kappa = {\cal N}{\rm
ln}4/a_0$\cite{Hwang2008},
where $a_0=\lambda_c/\alpha$ is the excitonic unit length.
Hence, the ratio of the Compton wavelength and screening length
$\kappa\lambda_c={\cal N}{\rm ln}4\alpha$ increases linearly with
$\alpha$  and the validity of the
quadratic approximation is
limited to  relatively weak interaction strengths $\alpha \lesssim({\cal
N}{\rm ln}4)^{-1}$. As can be recognized from the right panels of
Fig. \ref{dispersion_and_polarization},
for larger values of the coupling constant, relativistic modifications
to the dispersion enhance the effects of
screening thus reducing the gap.

In Fig. \ref{gap_vs_alpha}, we plot the quasiparticle gap as function of
coupling strength $\alpha$ for the full dispersion and
the quadratic approximation, respectively. Using the quadratic
approximation, the gap scales strictly with
the exciton energy unit, i.e. with $\alpha^2$, as shown in App. \ref{app:scales}.
Again, the comparison with the full solution shows that
the quadratic approximation yields reasonable results only for small
values of $\alpha$, while it overestimates the gap significantly
if the interaction strength increases.

\section{Conclusions}

In this paper, we present a selfconsistent approach to treat gap
generation and screening in an
effectively 2D excitonic insulator and applied our method to study the
groundstate properties of bilayer graphene.
To this end, we derive a generalization of the standard Lindhard formula
that is based on the equations of motion for the
density operator and is capable to describe the polarizability in a
nondiagonal representation.
To determine the groundstate properties, we solve the static limit of
the generalized Lindhard
formula selfconsistently with the gap equations, from which the
groundstate populations and excitonic coherences
are calculated.

Our method can be used on top of any effective single-particle band
structure calculation. For a weakly coupled system,
it reproduces the standard RPA result with self-energy corrections. In
the strongly coupled regime, spontaneously formed
coherent groundstate populations lead to the opening of a gap in the
quasi-particle spectrum, with a corresponding
suppression of the long-wavelength limit of the polarizability. This
mechanism leads to an enhancement of the
predicted gap.

For bilayer graphene, we find a gapped groundstate for any value of
the effecive fine structure constant.
Assuming a nominal value of $\alpha\approx 2.4$ for freely suspended BLG
in vacuum, we predict a gap of approximately
$14$ meV, which is within an experimentally accessible range. Increasing
(decreasing) the value of the coupling constant,
increases (decreases) the quasiparticle gap.
For small values of $\alpha$, we  find a
quadratic  dependence of the gap on
the effective coupling constant, that is found if a quadratic dispersion
is assumed\cite{Nandkishore2010}.
For large values of $\alpha\gtrsim 1/4{\rm ln}4$, the screening length is on the same order of or 
shorter than the Compton wavelength of the
quasiparticles. Linear corrections to the single-particle dispersion
make the simultaneous conservation of energy and momentum less
restrictive and thus enhance the effects of screening, leading to a subquadratic increase of the gap
with increasing coupling strength.

\appendix
\section{Derivation of the screened gap equations\label{app:SGE}}

\subsection{Screening in an effectively 2D system}

To derive the screened gap equations, we calculate the linear response
of the system to a perturbation by an
external scalar potential $V_{ext}(\br)$. The system Hamiltonian including the perturbation is given by
\begin{equation}
       H=H_0+H_C-e\int d^3r V_{ext}({\bm r})\Psi^\dagger(\bm r)\Psi(\bm r)
\end{equation}
where  $H_0$ is the single-particle part, $H_C$ describes
the electron-electron Coulomb interaction, and
\begin{equation}
     \rho({\bm r})= \langle \Psi^\dagger(\bm r)\Psi(\bm r)\rangle
\end{equation}
is the three-dimensional electronic density, respectively.
Assuming a system that is periodically with respect to the in-plane
coordinates,
the single-particle Hamiltonian
is given by
\begin{equation}
    H_0=\sum_{{\bk\in 1BZ}}\sum_{\alpha}\epsilon^0_{\bk\alpha}
c^\dagger_{\bk,\alpha}c_{\bk,\alpha}.
\end{equation}
where $\bk$ is a 2D wavevector and $\alpha$ denotes the band index.
The field operators can be expanded
in terms of the eigenfunctions of $H_0$; $\phi_{\bk\alpha}(\br)={\rm
exp}(i\bk\cdot\brho)u_{\bk\alpha}(\br)$
where
$u_{\bk\alpha}(\br)$ are the lattice-periodic Bloch waves:
\begin{equation}
      \Psi(\br)=\sum_{{\bk\in 1BZ}}\sum_{\alpha}c_{\bk\alpha}{\rm
e}^{i\bk\cdot\brho}u_{\bk\alpha}(\br).
\end{equation}
Here,   $\br=(\brho,z)$ is a three dimensional space coordinate.

By applying a coarse graining on the length scale of an elementary cell,
we obtain for the density operator
\begin{eqnarray}
       \hat\rho_{\bq}(z)&=&
       \sum_{{\bk\in 1BZ}}\sum_{\alpha\alpha'}W_{\alpha\alpha'}(\bk,\bq,z)
       c^\dagger_{\bk-\bq,\alpha}c_{\bk,\alpha'},
\end{eqnarray}
where $W_{\alpha\alpha'}(\bk,\bq,z)=
\int_{e.c.} d^2\rho\, u^*_{\bk-\bq,\alpha}(\br)u_{\bk,\alpha'}(\br)$ is
a weight factor
determined by the overlap of the Bloch waves with different band indices
and crystal momenta. The weight factor displays
the $z$-dependence of the electron density. For a quasi-2D sheet, this factor is
nonvanising only in a region strongly localized around the position of the layer.
The Coulomb interaction and the external perturbation can be expressed with
the aid of the coarse grained density operator as

\begin{eqnarray}\label{HC2}
  H_C
    &=&
      \frac{e^2}{2}\sum_{\bq\neq 0}\int\!\!\!\int
dz\,dz'\,\hat\rho_{\bq}(z)V_0(\bq,z-z')
      \hat\rho_{-\bq}(z'),
             \\
 H_{\rm ext}
     &=&
     -e\sum_{\bq}\int dz\, V_{ext}(\bq,z)\hat\rho_{-\bq}(z).
\end{eqnarray}
Here,
\begin{equation}
 V_0(\bq,z)=\frac{1}{NA}\frac{2\pi}{\epsilon_B q}{\rm e}^{-q|z|} 
\end{equation}
is the 2D Fourier transform of the bare Coulomb potential, $A$ is the
area of an elementary lattice cell, $N$ is the number elementary cells
included in the integration region, and $\epsilon_B$ is the constant
background dielectric constant that results from the substrate only.
In all following calculations, the limit $N\rightarrow\infty$ is
implicitly included.

For any particle conserving 2-point operator, we can
calculate the expectation value from the Heisenberg equation of motion
$i\hbar\partial_t\langle \hat O\rangle =\langle \left[\hat
O,H\right]\rangle $, giving
\begin{widetext}
\begin{eqnarray}
      i\hbar\frac{\rm d}{{\rm d} t}\langle 
c^\dagger_{\bk-\bq,\alpha}c_{\bk,\alpha'}\rangle
         &=&           
\left(\epsilon^0_{\bk,\alpha'}-\epsilon^0_{\bk-\bq,\alpha}\right)\left<c^\dagger_{\bk-\bq,\alpha}c_{\bk,\alpha'}\right>
            -e\sum_{\bq'}\int dz\,V_{ext}(\bq',z)
\left<
\left[c^\dagger_{\bk-\bq,\alpha}c_{\bk,\alpha'},\hat\rho_{-\bq'}(z)\right]\right>
                 \nonumber\\
             &+&\frac{e^2}{2}\sum_{\bq'\neq 0}\int \!\!\!\int dz\,
dz'\,V_0(\bq',z-z')
             \left<
    \left\{
                \hat\rho_{\bq'}(z),\left[c^\dagger_{\bk-\bq,\alpha}c_{\bk,\alpha'},\hat\rho_{-\bq'}(z')\right]\right\}
             \right>
\end{eqnarray}
\end{widetext}
where $\left\{\hat A,\hat B\right\}=\hat A\hat B+\hat B\hat A$ is the anticomutator.

To evaluate the expectation values,
we make the standard decomposition
\begin{eqnarray}
\label{HFfactorization}
   \langle \hat A\hat B\rangle  =
         \langle \hat A\rangle\langle \hat B\rangle+\langle \hat A\hat
B\rangle_X +
         \Delta\langle \hat A\hat B\rangle
\end{eqnarray}
into a direct term, an exchange term, and a correlation contribution.
Within the standard HF-approximation, all
correlation contributions are neglected. The remaining expectation
values are calculated within the random phase
approximation (RPA). Hence, it is assumed that
 $c_{\bk,\alpha}$ has a time dependence $\propto {\rm
e}^{-i\epsilon_{\bk,\alpha} t/\hbar}$.
Operator combinations with a phase that depends on a summation index
will average out in the sums such that only the contributions with a constant phase are kept. For the external source and the direct term 
in Eq. (\ref{EOMfull}), these are the contributions with  $\bq=\bq'$ only, and
\begin{widetext}
\begin{eqnarray}
    \left[c^\dagger_{\bk-\bq,\alpha}c_{\bk,\alpha'},\hat\rho_{-\bq}(z)\right]    &=&         
 \sum_{\beta}\left(W^*_{\beta\alpha'}(\bk,\bq,z)c^\dagger_{\bk-\bq,\alpha}c_{\bk-\bq,\beta}
         -W^*_{\alpha\beta}(\bk,\bq,z)c^\dagger_{\bk,\beta}c_{\bk,\alpha'}\right).
\end{eqnarray}
For the exchange terms, one has
\begin{eqnarray}
  \left<  \left\{
                \hat\rho_{\bq'}(z),\left[c^\dagger_{\bk-\bq,\alpha}c_{\bk,\alpha'},\hat\rho_{-\bq'}(z')\right]\right\}\right>_X
          =
              -{\sum_{\bk'\in
1BZ}}\sum_{\beta\beta'\gamma}W_{\beta\beta'}(\bk',\bq',z)
              \left(
               W_{\alpha'\gamma}(\bk-\bq',-\bq',z')\langle 
c^\dagger_{\bk'-\bq',\beta}c_{\bk-\bq',\gamma}\rangle\langle 
c^\dagger_{\bk-\bq,\alpha}c_{\bk',\beta'}\rangle
               \right.
                   \nonumber\\
               -\left. W_{\gamma\alpha}(\bk-\bq,-\bq',z')\langle 
c^\dagger_{\bk'-\bq',\beta}c_{\bk,\alpha'}\rangle\langle 
c^\dagger_{\bk-\bq+\bq',\gamma}c_{\bk',\beta'}\rangle
               \right)\nonumber\\
\end{eqnarray}
In the first term, only contributions with $\bk=\bk'$ survive the
phase averaging, while in the second term dominant contributions arise
from the
$\bk'=\bk-\bq+\bq'$ contributions, yielding
\begin{eqnarray}
 \left<  \left\{
                \hat\rho_{\bq'}(z),\left[c^\dagger_{\bk-\bq,\alpha}c_{\bk,\alpha'},\hat\rho_{-\bq'}(z')\right]\right\}\right>_X
%
         &\approx&
            \sum_\gamma\sum_{\beta\beta'}
            \left(              
W^*_{\gamma\alpha'}(\bk,\bq',z')W_{\beta\beta'}(\bk,\bq',z)\langle                
c^\dagger_{\bk-\bq',\beta}c_{\bk-\bq',\gamma}\rangle\langle 
c^\dagger_{\bk-\bq,\alpha}c_{\bk,\beta'}\rangle
            \right.
               \nonumber\\
        &-&
            \left.
                W^*_{\beta'\beta}(\bk-\bq,-\bq',z)
W_{\gamma\alpha}(\bk-\bq,-\bq',z')\langle 
c^\dagger_{\bk-\bq+\bq',\gamma}c_{\bk-\bq+\bq',\beta'}\rangle
                \langle  c^\dagger_{\bk-\bq,\beta}c_{\bk,\alpha'}\rangle
            \right).
\end{eqnarray}
Defining the renormalized single-particle energies
\begin{eqnarray}
       \epsilon_{\bk,\alpha\alpha'} &=&
             \delta_{\alpha\alpha'}\epsilon^0_{\bk,\alpha\alpha'}
             -e^2\sum_{\bq'\neq 0}\sum_{\beta\beta'}
             \int \!\!\!\int dz dz'\,V_0(\bq',z-z')
             W^*_{\beta\alpha}(\bk,\bq',z)W_{\beta'\alpha'}(\bk,\bq',z')
             \langle  c^\dagger_{\bk-\bq',\beta'}c_{\bk-\bq',\beta}\rangle
\end{eqnarray}
one finds within the RPA
\begin{eqnarray}
\label{EOMfull}
        i\hbar\frac{\rm d}{{\rm d} t}\langle 
c^\dagger_{\bk-\bq,\alpha}c_{\bk,\alpha'}\rangle
           &=&             
\sum_{\beta\beta'}\left(\epsilon_{\bk\alpha'\beta'}\delta_{\alpha\beta}           
 -\epsilon_{\bk-\bq,\beta\alpha}\delta_{\alpha'\beta'}\right)\langle 
              c^\dagger_{\bk-\bq,\beta}c_{\bk,\beta'}\rangle
             -e\int dz\,
  \left< \left[c^\dagger_{\bk-\bq,\alpha}c_{\bk,\alpha'},\hat\rho_{-\bq}(z)\right]    \right>
V_{s}(\bq,z)\nonumber\\
\end{eqnarray}
\end{widetext}
where
\begin{eqnarray}
     V_{s}(\bq,z)
         &=&
           \left(V_{ext}(\bq,z)-e\int dz'\,V_0(\bq,z-z')\langle
\hat\rho_{\bq}(z')\rangle\right)\nonumber\\
\end{eqnarray}
is the external potential screened by the induced charge density
$\langle \hat\rho_{\bq}(z')\rangle$.
Hence, the exchange  part of the Coulomb interaction leads to the energy
renormalization 
while the direct terms produces screening, i.e. it replaces the potential
of the bare perturbation by that of the
perturbation charge plus the induced charge density.

On the level of a mean-field theory, all observable quantities can be expressed in terms of the single-particle expectation values
$\langle  c^\dagger_{\bk-\bq,\alpha}c_{\bk,\alpha'}\rangle$ and the system dynamics is  completely determined by the E.O.M. \ref{EOMfull}.
For the single particle operators, these equations are equivalent to those derived from the effective mean-field Hamiltonian
\begin{eqnarray}
H_{MF}&=&\sum_{\alpha\alpha'}\sum_{\bk\in 1BZ}\epsilon_{\bk\alpha\alpha'}  c^\dagger_{\bk,\alpha}c_{\bk,\alpha'}\nonumber\\
&-&e\sum_{\bq}\int dzV_s(\bq,z)\hat\rho_{-\bq}(z) \, .
\end{eqnarray}
The system groundstate properties and excitation dynamics can be obtained from this Hamiltonian. Defining the Liouville operator
\begin{eqnarray}
     {\cal  L}(\bk,\bq)_{\alpha\alpha'\beta\beta'}&=&\delta_{\alpha\beta}\epsilon_{\bk,\alpha'\beta'}-\delta_{\alpha'\beta'}\epsilon_{\bk-\bq,\beta\alpha},
\end{eqnarray}
a formal integration of  the E.O.M. on the operator level gives the density-density response function
\begin{widetext}
\begin{eqnarray}
   \Pi(\omega,\bq,z,z') &=&
-\frac{i}{\hbar}\int_0^\infty dt{\rm e}^{i\omega t}\left.\left<\left[\hat\rho_{\bq}(z,t),\hat\rho_{-\bq}(z')\right]\right> \right|_{V_{ext}=0}
\nonumber\\
&=&
             \sum_{\alpha\alpha'\beta\beta'} \sum_{\bk\in 1BZ}{ W}_{\alpha\alpha'}(\bk,\bq,z)
\left(\hbar\omega {\cal E}-{\cal
L}(\bk,\bq)\right)_{\alpha\alpha'\beta\beta'}^{-1}\left.\left<\left[c_{\bk-\bq,\beta}^\dagger c_{\bk,\beta'},\hat\rho_{-\bq}(z')\right]\right>\right|_{V_{ext}=0}.
\label{app:densdensresponse}
\end{eqnarray}
\end{widetext}
In the presence of an external perturbation, 
we obtain the set of equations
\begin{eqnarray}
\label{rhocg1}
        \langle \hat\rho_{\bq}(z)\rangle &=&
             -e\int dz'\Pi_0(\omega,\bq,z,z')V_s(\bq,z')\\
         V_s(\bq,z) &=&
              V_{ext}(\bq,z)-e\int dz'\,V_0(\bq,z-z')\langle
\hat\rho_{\bq}(z')\rangle.\nonumber\\
\end{eqnarray}
Provided the polarization function is known, this set of equations  can
in principle be solved
iteratively. 

As can be recognized from Eq. \ref{app:densdensresponse}, the
$z,z'$-dependence of the polarization function is entirely
determined by the weight functions $W_{\alpha\alpha'}(\bk,\bq,z)$, i.e.
by the eigenfunctions of the noninteracting problem.
For a general situation with arbitrary Bloch waves, the resulting
Dyson-like series cannot be summed into a closed
analytical expression due to the nontrivial $z$-dependence.
As this is essential for the division into groundstate and dynamical
parts, in the following we shall restrict
to the TB approximation including only a single atomic orbital.

Within the TB approximation, we  make the  ansatz
\begin{eqnarray}
     u_{\bk\alpha}(\br) &=&
            \sum_{i}\sum_{{\bR}_i}C^i_\alpha(\bk){\rm
e}^{i\bk\cdot(\br-{\bR}_i)}\phi(\br-{\bR}_i)
\end{eqnarray}
where the $i$-sum runs over the different sublattices. Both the
z-dependence of the charge density and the polarization
function are determined by the weight factor, for which we find
within the nearest neighbor approximation
\begin{eqnarray}
        W_{\alpha\alpha'}(\bk,\bq,z) &=&
           \sum_i C^{i*}_{\alpha}(\bk-\bq)C^i_{\alpha'}(\bk)f(\bq,z-z_i ).
\end{eqnarray}
Here,
\begin{eqnarray}
      f(\bq ,z)&=&\int d^2\rho \,{\rm e}^{i\bq\cdot{\bm\rho}}|\phi(\bm r)|^2
\end{eqnarray}
is the in-plane Fourier transform of the atomic electron density.
Hence, within the TB approximation, the $z$ dependence decouples from
the  band indices, and it is easily verified that
\begin{eqnarray}
     \langle \hat\rho_{\bq}(z)\rangle &=&
             \sum_i\rho_{\bq}^{i}f(\bq,z-z_i)\\
     \Pi_0(\omega,\bq,z,z') &=&
             \sum_{ij}f(\bq,z-z_i)\Pi_0^{ij}(\omega,\bq)f(-\bq,
z'-z_j),\nonumber\\
\end{eqnarray}
where $\rho^i_{\bq}$ is the electron density localized on the sublattice $i$
and $\Pi_0^{ij}(\omega,\bq)$ is the density-density
response function between sublattices $i$ and $j$, respectively. Here, we defined
\begin{eqnarray}
      V_S^j(\bq)&:=&\int dz\,f(-\bq,z-z_j)V_S(\bq,z)\\
      V_0^{ij}(\bq)&:=& \int\!\!\!\int dz
dz'\,f(-\bq,z-z_i)V_0(\bq,z-z')f(\bq,z-z_j)\nonumber\\
                   &=& V_0(\bq)H(qd,|z_i-z_j|),
\end{eqnarray}
 $V_0^{ij}(\bq)$ is the bare Coulomb potential between carriers in sheet
$i$ and $j$,
and
\begin{eqnarray}
       H(qd,L)&=&\int dz\int dz'{\rm e}^{-q|z-z'+L|}f(\bq,z)f(-\bq,z')
\nonumber\\
\end{eqnarray}
accounts for the localization.
Using these definitions, the $z$-integrals reduce to a matrix
multiplication and we find
\begin{eqnarray}
\label{def:epsmatrix}
       {\underline{V}_S}(\omega,\bq) &=&
             \left(\underline{\underline{I\!\!
I}}-e^2{\underline{\underline{V}}}_0(\bq)           
 {\underline{\underline{\Pi}}}(\omega,\bq)\right)^{-1}{\underline{V}}_{ext}(\bq)\nonumber\\
          &\equiv&           
 {\underline{\underline{\epsilon}}}(\omega,\bq)^{-1}{\underline{V}}_{ext}(\bq)
\end{eqnarray}
with the matrix-valued dynamical dielectric  function
${\underline{\underline{\epsilon}}}(\omega,\bq)$.
If all atoms are located at the same $z$-positions, Eq.
\ref{def:epsmatrix} simplifies to
\begin{eqnarray}
     {\underline{V}_S}(\omega,\bq) &=&
              \epsilon^{2D}(\omega,\bq)^{-1}{\underline{V}}_{ext}(\bq)
\end{eqnarray}
with
\begin{eqnarray}
\label{def:epsscalar}
        \epsilon^{2D}(\omega,\bq) &=&
            1-e^2V_0(\bq)H(qd,0)\sum_{ij}\Pi^{ij}(\omega,\bq).
\end{eqnarray}

\subsection{Screened gap equations}

The expression derived for the dynamical dielectric function
${\underline{\underline{\epsilon}}}(\omega,\bq)$
and its static limit depend on the
expectation values $\langle 
c^\dagger_{\bk,\alpha}c_{\bk,\alpha'}\rangle$ via the renormalized
single-particle energies
and the source terms. To calculate these, we use the E.O.M. \ref{EOMfull}
with $\bq=0$ with the screened Coulomb potential
in the absence of an external perturbation. For simplicity, we will
restrict our analysis to a two-band model, and refer to the bands 
as conduction and valence band whose band indices are denoted by $s=\pm$, respectively. Using the variables
$f_{\bk,s}=\langle  c^\dagger_{\bk,s}c_{\bk,s}\rangle$ and
$P_{\bk}=\langle  c^\dagger_{\bk,-}c_{\bk,+}\rangle$
and defining
\begin{eqnarray}
     \Sigma_{\bk,s}&=& \epsilon_{\bk,s,s}\\
     \Omega_{\bk}&=&-\epsilon_{\bk,+,-}
\end{eqnarray}
we find the standard semiconductor Bloch equations (SBE) \cite{Lindberg1988}
and the conservation law
\begin{eqnarray}\label{SBE}
     i\hbar\partial_t P_{\bk}(t) &=&       
 \left(\Sigma_{\bk,+}-\Sigma_{\bk,-}\right)P_{\bk}+\left(f_{\bk,+}-f_{\bk,-}\right)\Omega_{\bk},
             \nonumber\\
     i\hbar\partial_t f_{\bk,s}(t) &=&
          s\left(\Omega^*_{\bk} P_{\bk}-\Omega_{\bk} P^*_{\bk}\right),
          \nonumber\\
      0 &=&           
\partial_t\left(|P_{\bk}|^2+\frac{1}{4}(f_{\bk,+}-f_{\bk,-})^2\right),
\end{eqnarray}
that must be solved with the appropriate initial and boundary conditions.
For the groundstate, we require that all dynamical variables are
stationary and adiabatically connected to the noninteracting
groundstate with $f_{\bk} \equiv f_{\bk,+} = 1-f_{\bk,-}=0$ and
$P_{\bk}=0$,
giving the algebraic relations
\begin{eqnarray} 
 P_{\bk}&=&\frac{\Omega_{\bk}}{2\sqrt{\Sigma_{\bk}^2+|\Omega_{\bk}|^2}},\label{defP_app}\\ 
 f_{\bk}&=&\frac{1}{2}\left(1-\frac{\Sigma_{\bk}}{\sqrt{\Sigma_{\bk}^2+|\Omega_{\bk}|^2}}
\right)\label{deff_app}
\end{eqnarray}
with $\Sigma_\bk=(\Sigma_{\bk,+}-\Sigma_{\bk,-})/2$.

Inserting this into the definitions for the  renormalized energies gives
the gap equations:
\begin{widetext}
\begin{eqnarray}
    \Sigma_{\bk} &=&
            \frac{1}{2}\left(\epsilon^0_{\bk,+}-\epsilon^0_{\bk,-}\right)
           -e^2\sum_{\bq\neq
0}\left(V_{-+-+}(\bk,\bq)-V_{----}(\bk,\bq)\right)\nonumber\\
            &-&\frac{e^2}{2}\sum_{\bq\neq 0}           
\left(V_{++++}(\bk,\bq)+V_{----}(\bk,\bq)-V_{-+-+}(\bk,\bq)-V_{+-+-}(\bk,\bq)\right)           
\left(1-\frac{\Sigma_{\bk-\bq}}{\sqrt{\Sigma_{\bk-\bq}^2+|\Omega_{\bk-\bq}|^2}}\right)\nonumber\\
            &-&e^2\sum_{\bq\neq
0}\Re\left[\left(V_{++-+}(\bk,\bq)-V_{+---}(\bk,\bq)\right)
           \frac{\Omega_{\bk-\bq}}{\sqrt{\Sigma_{\bk-\bq}^2+|\Omega_{\bk-\bq}|^2}}\right]\label{gap:sigma}
                       \\
   \Omega_{\bk} &=&
            e^2\sum_{\bq\neq 0}V_{-+--}(\bk,\bq)\nonumber\\
            &+&\frac{e^2}{2}\sum_{\bq\neq 0}
            \left(V_{+++-}(\bk,\bq)-V_{-+--}(\bk,\bq)\right)
           \left(1-\frac{\Sigma_{\bk-\bq}}{\sqrt{\Sigma_{\bk-\bq}^2+|\Omega_{\bk-\bq}|^2}}\right)\nonumber\\
            &+&\frac{e^2}{2}\sum_{\bq\neq 0}
           \left(V_{++--}(\bk,\bq)\frac{\Omega_{\bk-\bq}}{\sqrt{\Sigma_{\bk-\bq}^2+|\Omega_{\bk-\bq}|^2}}+
           V_{-++-}(\bk,\bq)\frac{\Omega^*_{\bk-\bq}}{\sqrt{\Sigma_{\bk-\bq}^2+|\Omega_{\bk-\bq}|^2}}
            \right)
\label{gap:omega}
\end{eqnarray}
where the Coulomb matrix elements
\begin{eqnarray}
        V_{ss's''s'''}(\bk,\bq)=\int \!\!\!\int dz dz'\,
W^*_{ss'}(\bk,\bq,z)V_s(\bq,z-z')W_{s''s'''}(\bk,\bq,z'),
\end{eqnarray}
\end{widetext}
 have to be evaluated selfconsistently with the screened Coulomb
potential. In Eqs.\ref{gap:sigma} and \ref{gap:omega} the constant
contributions in the first lines are the energy and gap renormalizations
resulting from the filled valence bands. The contributions have to be dropped if band parameters are used that have been taken from
experimental values or band structure calculations that include (parts)
of the e-e-interactions in the filled valence band.

\section{Derivation of the model Hamiltonian\label{app:Hamiltonian}}

Here, we derive the two-band model Hamiltonian
starting  from Eq.\ref{bilayer-h} given in Sec. \ref{sec:Hamiltonian}.
Diagonalization of Eq.\ref{bilayer-h} gives four bands that are parabolic
in the vicinity of the two Dirac points:
\begin{equation}
     H_0=\sum_{s=\pm,\sigma=\pm}\sum_{\bk} s
E^\sigma_{\bk}c_{s,\sigma,\bk}^\dagger c_{s,\sigma,\bk}
\end{equation}
with
\begin{eqnarray}
    E^\sigma_{\bk}&=&\sqrt{\frac{t_1({\bk})+\sigma \sqrt{t_2({\bk})}}{2}}\\
    t_1({\bk})&=&\gamma_1^2+(2\gamma_0^2+\gamma_2^2)|f({\bk})|^2\\   
t_2({\bk})&=&(\gamma_1^2-\gamma_2^2|f({\bk})|^2)^2+4\gamma_0^2|f({\bk})|^2|\gamma_1+\gamma_2f({\bk})|^2.\nonumber\\
\end{eqnarray}

At the Dirac points, the $E_{-,\bk}$ bands are degenerate, while the
$E_{+,{\bk}}$ are
shifted by $\pm\gamma_1$. Close to the Dirac points, $\gamma_2
f({\bm K}^\pm+{\bk})=-\frac{3\gamma_2b}{2}e^{-i\pi/6}(k_x\pm ik_y)=-
\frac{\gamma_2}{\gamma_0}\hbar v_F k\,{\rm exp}i(\phi_{\bk}\mp\pi/6)$,
introducing an anisotropy into the bandstructure.
However, while contributions arising from the $B1$-$A2$ coupling
$\propto\gamma_1$ are independent of $\bk$, the
dominant terms $\propto\gamma_2$ are linear in $k$ and $\propto
\gamma_2/\gamma_0\approx 0.1$. Hence, close to
the Dirac points, the trigonal warping can be neglected. Putting
$\gamma_2=0$,
the dispersion is simplified to
\begin{eqnarray}
\label{E_wowarping}
     E^\pm_{\bK^\pm+\bk} &=&             
 \sqrt{\left(\frac{\gamma_1}{2}\right)^2+\gamma_0^2|f(\bK^\pm+\bk)|^2}\pm\frac{\gamma_1}{2}\nonumber\\
                         &=&
               \sqrt{\left(\frac{\gamma_1}{2}\right)^2+(\hbar v_F
k)^2}\pm\frac{\gamma_1}{2}.
\end{eqnarray}
In the following, all considerations will be restricted to the
low-energy regime. 

We use the valley index $\tau$ to define
$f_\tau(\bk)=f(\bK^\tau+\bk)$ and
the operators
$c_{\tau,s,\sigma,\bk}=c_{s,\sigma,\bK^\tau+\bk}$ that annihilate an
electron with band index $s,\sigma$ and momentum $\bK^\tau+\bk$. With
this abbreviation,
the band operators are related to the lattice operators by the unitary
transformation
\begin{eqnarray}
    \begin{pmatrix}
        c_{\tau,+,+,\bk}\cr c_{\tau,+,-,\bk}\cr c_{\tau,-,-,\bk}\cr
c_{\tau,-,+,\bk}
    \end{pmatrix}
           &=&
           U_\tau
              \begin{pmatrix}
                  a_{1,\bK^\tau+\bk}\cr b_{1,\bK^\tau+\bk}\cr
a_{2,\bK^\tau+\bk}\cr b_{2,\bK^\tau+\bk}
              \end{pmatrix}.\\
\end{eqnarray}
\begin{widetext}
\begin{eqnarray}
   U_\tau &=&
          \frac{1}{\sqrt{2(E^+_{\bk}+E^-_{\bk})}}
           \begin{pmatrix}
                  \displaystyle{\sqrt{E^+_{\bk}}}
&-\displaystyle{\frac{\gamma_0f_\tau^*(\bk)}                 
{\sqrt{E^+_{\bk}}}}&\displaystyle{\frac{\gamma_0f_\tau(\bk)}{\sqrt{E^+_{\bk}}}}
                  &\displaystyle{\sqrt{E^+_{\bk}}}
                                               \cr
                  \displaystyle{-\sqrt{E^-_{\bk}}}
&-\displaystyle{\frac{\gamma_0f_\tau^*(\bk)}
                 {\sqrt{E^-_{\bk}}}}&\displaystyle{\frac{\gamma_0f_\tau(\bk)}{\sqrt{E^-_{\bk}}}}
                  &\displaystyle{\sqrt{E^-_{\bk}}}
                                                \cr
                  \displaystyle{\sqrt{E^-_{\bk}}}
&-\displaystyle{\frac{\gamma_0f_\tau^*(\bk)}
                 {\sqrt{E^-_{\bk}}}}&\displaystyle{\frac{\gamma_0f_\tau(\bk)}{\sqrt{E^-_{\bk}}}}
                  &\displaystyle{\sqrt{E^-_{\bk}}}
                                                 \cr
                  \displaystyle{-\sqrt{E^+_{\bk}}}
&\displaystyle{\frac{\gamma_0f_\tau^*(\bk)}
                 {\sqrt{E^+_{\bk}}}}&-\displaystyle{\frac{\gamma_0f_\tau(\bk)}{\sqrt{E^+_{\bk}}}}
                  &\displaystyle{\sqrt{E^+_{\bk}}}
                                                 \cr
           \end{pmatrix}
\end{eqnarray}
Close to the Dirac point, if $\hbar v_F k<<\gamma_1$, the unitary
transformation can be approximated by
\begin{eqnarray}
      U_\tau =
           \frac{1}{\sqrt{2}}
           \begin{pmatrix}
             
 \displaystyle{1}&\displaystyle{0}&\displaystyle{0}&\displaystyle{1}
                             \cr
               \displaystyle{0}&-\displaystyle{{\rm
e}^{-i(\tau\phi_{\bk}-\pi/6)}}
               &\displaystyle{{\rm
e}^{i(\tau\phi_{\bk}-\pi/6)}}&\displaystyle{0}
                             \cr
               \displaystyle{0}&-\displaystyle{{\rm
e}^{-i(\tau\phi_{\bk}-\pi/6)}}
               &\displaystyle{{\rm
e}^{i(\tau\phi_{\bk}-\pi/6)}}&\displaystyle{0}
                             \cr
             
 -\displaystyle{1}&\displaystyle{0}&\displaystyle{0}&\displaystyle{1}
           \end{pmatrix}
           +\frac{\hbar v_F k}{\gamma_1}\frac{1}{\sqrt{2}}
           \begin{pmatrix}
            \displaystyle{0}&-\displaystyle{{\rm
e}^{-i(\tau\phi_{\bk}-\pi/6)}}
            &\displaystyle{{\rm
e}^{i(\tau\phi_{\bk}-\pi/6)}}&\displaystyle{0}
                                \cr
           
 -\displaystyle{1}&\displaystyle{0}&\displaystyle{0}&\displaystyle{1}
                                \cr
             
\displaystyle{1}&\displaystyle{0}&\displaystyle{0}&\displaystyle{1}
                                \cr
             -\displaystyle{0}&\displaystyle{{\rm
e}^{-i(\tau\phi_{\bk}-\pi/6)}}
             &-\displaystyle{{\rm
e}^{i(\tau\phi_{\bk}-\pi/6)}}&\displaystyle{0}
           \end{pmatrix}
                          \nonumber\\
                +O\left(\left(\frac{\hbar v_F k}{\gamma_1}\right)^2\right).
\end{eqnarray}
\end{widetext}
Hence, in the vicinity of the Dirac points, the upper conduction and
lower valence band consist of the
uncoupled sublattices components $A1$ and $B2$,
while the lower, degenerate bands contain the interacting $A2$
and $B1$ components.
Within this range,
$E^-_{\bk}<(\sqrt{5}-1)\gamma_1/2=0.62\gamma_2<E^+_{\bk=0}$, the
upper bands can be neglected and the Bloch functions of the lower bands are
approximately given by
\begin{eqnarray}
       u_{\tau,s,-,\bk}(\br) &=&
                -\frac{{\rm e}^{-i\bk\cdot\brho}}{\sqrt{2}}\left({\rm
e}^{-i(\tau\phi_{\bk}-\pi/6)}
                \psi_{B1,\bk}(\br)\right.\nonumber\\
                &+&\left. s{\rm
e}^{i(\tau\phi_{\bk}-\pi/6)}\psi_{A2,\bk}(\br)\right)\\
       \psi_{B1,\bk}(\br) &=&
                 \frac{1}{\sqrt{N}}\sum_{\bR_B^1}{\rm
e}^{i\bk\cdot\bR_B^1}\phi(\br-\bR_B^1).\\
        \psi_{A2,\bk}(\br) &=&
                 \frac{1}{\sqrt{N}}\sum_{\bR_A^2}{\rm
e}^{i\bk\cdot\bR_A^2}\phi(\br-\bR_A^2).
\end{eqnarray}
Taking into account only on-site contributions, one finds for the weight
functions
\begin{eqnarray}
\label{weight}
      W_{\tau,ss'}(\bk,\bq,z) &=&
                  \frac{1}{2}\left({\rm
e}^{-i(\tau\phi_{\bk}-\phi_{\bk-\bq})}f({\bq},z-z_1)\right.\nonumber\\
                  &+&\left. ss'{\rm
e}^{i(\tau\phi_{\bk}-\phi_{\bk-\bq})}f({\bq},z-z_2)\right)
\end{eqnarray}
with
\begin{eqnarray}
      f(\bq,z)&=&\int_{e.c.} d^2\rho {\rm e}^{-i\bq\cdot\brho}|\phi(\br)|^2.
\end{eqnarray}
The weight functions consist of parts localized in a particular sheet, such that
the total charge density can be divided into the respective contributions $\rho^i$,
located in the layer $i$.
Inserting this into the Coulomb Hamiltonian, one finds
\begin{eqnarray}
   H^{2B}_C &=&         
\frac{e^2}{2}\sum_{\bq}\sum_{ij}V^{ij}_0(\bq)\hat\rho^{i}_{\bq}\hat\rho^{j}_{-\bq}
\end{eqnarray}
where
\begin{eqnarray}
        V_0^{ij}(\bq) &=&  V_0(\bq)\iint dzdz' f(\bq,z-z_i){\rm
e}^{-q|z-z'|}f(-\bq,z'-z_j)\nonumber\\
                      &=& V_0(\bq)H(q,|z_i-z_j|)
\end{eqnarray}
is the Coulomb potential between carriers in the layers $i$ and $j$
and
\begin{eqnarray}
       H(q,L)&=&\int dz\int dz'{\rm e}^{-q|z-z'+L|}f(\bq,z)f(-\bq,z')
\end{eqnarray}
accounts for the localization.

\section{Involved energy and length scales\label{app:scales}}

Dealing with excitonic properties requires the distinction between an
atomic and excitonic length scale,
where the excitonic length is  assumed to be large compared to
the atomic length scale in order to describe collective properties
within the continuum approximation.
Lengths that enter on the atomic scale are
the effective thickness $d$ of a single graphene sheet, the
carbon-carbon distance $b$ and the interlayer spacing $L$.
As a practical definition for the effective sheet thickness, one can use
the scaling length of the atomic orbitals constituting the valence and
conduction band:
$\phi(\br)=d^{-3/2}\tilde\phi(\br/d)$.
Within a certain range, the interlayer spacing $L$ can be considered as
independent from the sheet thickness.
However, in order for the layers to be electronically coupled, the
interlayer spacing has to be on the same order of
magnitude as the layer thickness.
Both, the layer thickness and interlayer spacing enter as intrinsic
lengths into the Coulomb interaction:
\begin{eqnarray}
   V^{ij}(\bq) &=& V_0(\bq)H(q,|z_i-z_j|)=V_0(\bq)\tilde
H(qd,|z_i-z_j|/d),\nonumber\\
\end{eqnarray}
where $\tilde H(qd,|z_i-z_j|/d)$ depends on the scaled, dimensionless
quantities $qd$ and $L/d$ only. In the region
$qd\sim qL\ll 1$, $\tilde H(qd,|z_i-z_j|/d)\approx 1$ independently of the
scaled interlayer distance $L/d$.

Using the full relativistic dispersion \ref{E_wowarping}, the intrinsic
length for the kinetic energy is
the de Broglie wavelength $\lambda_c=\hbar/mv_F$, where the mass is related
to the hopping parameters $\gamma_0$ and $\gamma_1$ and the carbon-carbon
distance $b$ via $\hbar^2/2m=9\gamma_0^2b^2/4\gamma_1$. On a length
scale larger than the de Broglie wavelength,
relativistic effects may be neglected and the dispersion can be
considered to be quadratic, while it becomes
linear on shorter length scales.

Collective excitonic properties emerge from the interplay between the
kinetic and Coulombic energies.
Defining the scaled quantities
$\tilde c_{s,\bx}= c_{s,\bx/\lambda}$,
$\tilde\rho_{\bx}=\hat\rho_{\bx/\lambda}$,
$\tilde V_{0,\lambda}^{ij}(\bx)=\lambda V_0^{ij}(\bx/\lambda)$,
one finds within the two-band continuum approximation
\begin{eqnarray}
  H_\lambda &=&
        \frac{\hbar v_F}{\lambda_c}\left(\sum_{ s,\bx} s
\left(\sqrt{1+\left(\frac{\lambda_c}{\lambda}x\right)^2}-1\right)
         \tilde c_{s,\bx}^\dagger \tilde c_{s,\bx}\right.\nonumber\\       
 &+&\left.\alpha\frac{\lambda_c}{\lambda}\sum_{\bx}\sum_{ij}\tilde
V_{0,\lambda}^{ij}(\bx)\tilde\rho^{i}_{\bx}
         \tilde\rho^{j}_{-\bx}\right).
\end{eqnarray}
Choosing $\lambda=\lambda_c/\alpha=a_0=\hbar^2\epsilon_{\rm b}/me^2$,
the scaled Hamiltonian becomes independent of
$\lambda$ if
\begin{equation}
\tilde V_{0,a_0}^{ij}(\bx)=a_0 V_0(\bx/a_0)\tilde
H(xd/a_0,|z_i-z_j|/d)\equiv \tilde V_{0}^{ij}(\bx)
\label{scale:Coulomb}
\end{equation}
is independent  of $a_0$ within the relevant $x$-range.
In the strict 2D limit, $\tilde H(qd,L/d)\equiv 1$ and the above
condition is fulfilled for the bare Coulomb
potential $V_0(\bq)=2\pi/NAq$.
If the exciton Bohr radius is large compared to the effective thickness,
i.e.
if $d/a_0=\alpha(\gamma_1/3\gamma_0)(d/b)\ll1$, finite size effects
become negligible and the system can be considered to be effectively 2D.
 Here, the ratio $d/b$ is fixed by the carbonic wave function making the
$\sigma$-bonds and  valence band respectively, $\gamma_1/\gamma_0$ is
fixed by the hopping parameters. Though the scaled Hamiltonian is
independent of the scaling length, it still depends on the effective
fine-structure
constant $\alpha=e^2/\epsilon_{\rm B}\hbar v_F$ that can be varied
within a certain range by changing the screening properties of the
dielectric environment.
Since the Bohr radius increases inversely with decreasing $\alpha$, the
2D approximation will be particularly good for small values of the
effective
coupling. 

In general, the fine-structure constant is a measure for the
importance of relativistic effects. In the nonrelativistic region
 $\alpha<1$, the scaled single particle dispersion
$(\sqrt{1+\alpha^2x^2}-1)\approx \alpha^2 x^2/2$ is
approximately quadratic, and  the total  Hamiltonian is proportional to
$\alpha^2$. Hence, within the nonrelativistic
 2D limit, all lengths scale strictly with the exciton Bohr radius and
all energies with the exciton energy unit
$E_0=me^4/\epsilon_{\rm B}^2\hbar^2=\alpha^2\gamma_1/2$, respectively. This property that
should be conserved within any additional level of
approximation.
Particularly, in the nonrelativistic limit, Eq. \ref{scale:Coulomb} is not only valid for the bare,
but also for the screened Coulomb potential and one obtains the scaled
gap equations
\begin{eqnarray}
     \tilde\Omega_{\bx} &=&
               \sum_{\bx'}\tilde
V(|\bx-\bx'|)\frac{\tilde\Omega_{\bx'}}{E_{\bk'}}
\label{scaled:gapomega}
\\
     \tilde\Sigma_{\bx} &=&
             \tilde\epsilon_{\bx}+
             \sum_{\bx'}\tilde V(|\bx-\bx'|){\rm
cos}2(\phi_{\bx}-\phi_{\bx'})\frac{\tilde\Sigma_{\bx'}}{E_{\bk'}}
\label{scaled:gapsigma}
\end{eqnarray}
with $\Omega_{\bk}=E_0\tilde\Omega_{a_0\bk}$ and
$\Sigma_{\bk}=E_0\tilde\Sigma_{a_0\bk}$.
The scaled gap equations must be solved with the screened potential
\[\tilde V(\bx)=\frac{\tilde V_0(\bx)}{1-e^2\tilde
V_0(\bx)\tilde\Pi^{2D}(\tilde\omega=0,\bx)},\]
and
$\tilde\Pi^{2D}(\tilde\omega,\bx)=E_0\Pi^{2D}(\omega/E_0,\bx/a_0)$ is the
scaled polarization function
\begin{widetext}
\begin{eqnarray}
  \tilde \Pi(\tilde\omega=0,\bx) &=&
             -\sum_{ss'\sigma\sigma'} \sum_{\bx'\in 1\tilde BZ}{\tilde  W}_{ss'}(\bx',\bx)
\tilde{\cal
L}(\bx',\bx)_{ss'\sigma\sigma'}^{-1}\left.\left<\left[\tilde c_{\bx'-\bx,\sigma}^\dagger\tilde c_{\bx',\sigma'},\hat{\tilde\rho}_{-\bx}\right]\right>\right|_{V_{ext}=0}.
\end{eqnarray}

\end{widetext}


\raggedright

%

\end{document}